\def\lsi{\raise0.3ex\hbox{$<$\kern-0.75em\raise-1.1ex\hbox{$\sim$}}}
\def\gsi{\raise0.3ex\hbox{$>$\kern-0.75em\raise-1.1ex\hbox{$\sim$}}}
\newcommand{\lsim}{\mathop{\lsi}}
\newcommand{\gsim}{\mathop{\gsi}}
\newcommand\threev[1]{{\bf #1}}
\newcommand\twov[1]{\vec{#1}}

\newcommand\PRD{{\em Phys.\ Rev.\ D\ }}
\newcommand\PRL{{\em Phys.\ Rev.\ Lett.\ }}

\newcommand\MPLA{{\em Mod.\ Phys.\ Lett.\ A\ }}

\newcommand\Nat{{\em Nature\ }}
\documentclass[prd%
,aps%
,twocolumn%
,amssymb%
,preprintnumbers%
,nofootinbib%
]{revtex4}

\begin{document}

\title{Estimation of vortex density after
superconducting film quench}

\author{T.W.B.~Kibble}
\affiliation{Blackett Laboratory, Imperial
College London,\\London SW7 2AZ, United Kingdom}

\author{A.~Rajantie}
\affiliation{DAMTP, CMS, University of Cambridge,
Wilberforce Road,\\
Cambridge CB3 0WA, United Kingdom
}

\date{25 June, 2003}

\begin{abstract}
This paper addresses the problem of vortex formation
during a rapid quench in a superconducting film.  It
builds on previous work showing that in a local gauge
theory there are two distinct mechanisms of defect
formation, based on fluctuations of the scalar and
gauge fields, respectively.  
We show how vortex
formation in a thin film differs from the fully two-dimensional case,
on which most theoretical studies have focused. We discuss ways of
testing theoretical predictions in superconductor experiments
and analyse the results of recent experiments in this light.
\end{abstract}

\preprint{DAMTP-2003-60, Imperial/TP/02-03/26}
\preprint{cond-mat/0306633}
\maketitle

\section{Introduction}

There has been much recent interest in testing
cosmological defect-formation scenarios
\cite{Kib80,Zur96,HinR00} in the laboratory.  In the Nineties, 
experiments in
liquid crystals \cite{Chu+91,Bow+94,Digal:1998ak} and
especially in liquid $^3$He \cite{Bau+96,Ruu+96}
clearly showed the formation of defects in line with
predictions.  In $^4$He, experiments have proved negative
\cite{Dod+98}, but there is good reason to believe
that this is due to the fact that vortices disappear
too rapidly to be seen \cite{Riv00,Hen+00}.  

More recently, the focus of the experimental activity has shifted
towards
superconductors~\cite{CarP99,Car+00,Kav+00,Mon+02,Mon202,Kir03,Man03}.
In the pioneering experiment by Carmi and Polturak~\cite{CarP99},
no evidence of spontaneous vortex formation was seen, but
later experiments with a multiple-Josephson-junction
loop \cite{Car+00} and with an annular Josephson
junction \cite{Kav+00,Mon+02,Mon202} have been positive.
Earlier this year, successful experiments have also been carried out
with an array of superconducting rings~\cite{Kir03} and a thin
superconducting film~\cite{Man03}.

It is not possible to apply the original estimates of
defect numbers directly to the case of the
superconducting transition, because it involves a
local gauge theory rather than a global symmetry
breaking \cite{RudS93,CopS96,HinR00}.  In
ref.~\cite{CarP99}, a somewhat \emph{ad hoc}
modification of the standard prediction was proposed
for comparison with experiment, but a more firmly
based estimate is needed.  The problem of defect
formation in a gauge theory has been analyzed by
Hindmarsh and Rajantie \cite{HinR00}, who
showed that in addition to
the standard mechanism of defect formation, there is a
second mechanism operating. This flux trapping mechanism stems from the
fluctuations in the magnetic field, which get frozen out during the quench.

The two mechanisms produce defect distributions with
characteristically very different correlation
properties.  In a gauge theory, one would expect positive
vortex-vortex correlations at short distances. In other words, the
vortices should be formed in clusters of equal sign.
This is supported by numerical simulations carried out in the Abelian
Higgs model~\cite{HinR01,Stephens:2001fv}. On the other hand, the
vortices should have negative correlations at all distances in global
theories. 

One of us \cite{Raj03} has studied an
SO(3) gauge theory, and shown that monopoles form,
even when there is no true phase transition, only a
smooth crossover.  The monopole distribution again
shows positive monopole-monopole correlation, in
contrast to the situation in a global symmetry theory,
where one would expect monopole-antimonopole
correlations.

In this paper, we examine the spontaneous formation of
vortices (or fluxons or fluxoids) in a thin film undergoing the
normal-to-superconducting transition in the light of existing and
future experiments.
This setup
differs significantly from the fully two-dimensional
case studied in Ref.~\cite{Stephens:2001fv},  because
the magnetic field extends outside the film and leads
to long-range
$1/r$ interaction between vortices~\cite{Pea64}. Therefore, there
are actually many similarities with monopole formation
in three dimensions. In both cases, the interaction is
screened by thermal effects, although the screening is
less effective here.

We also discuss the results of the experiments carried out so far, 
and conclude that whilst they are compatible with the two mechanisms,
they do not currently constitute a positive confirmation of either.

Throughout this paper, we use the natural units $c=\hbar=k_B=1$.
We will write three-vectors with bold-faced symbols as $\threev{k}$, and
two-vectors on the film with an arrow as $\twov{k}$.

\section{Two-dimensional superconductors}
\label{sect:2d}

Theoretical studies of defect formation in two
dimensions have shown that there are essentially two
different regimes. At a weak enough coupling or low
enough critical temperature, the relevant degree of
freedom is the order parameter field and the process
is well described by the Kibble-Zurek mechanism~\cite{Kib80,Zur96}. 
On
the other hand, at strong coupling or high critical
temperature, it is dominated by thermal fluctuations
of the gauge field~\cite{HinR00}.

The simplest example of the Kibble-Zurek 
mechanism is
the phase transition in a Ginzburg-Landau model with a
global U(1) symmetry. 
Because the mechanism is very generic, we can consider a two-dimensional
relativistic version of the model, with the
Lagrangian density
\begin{equation}
{\cal L}=\dot\phi^*\dot\phi-\vec{\nabla}\phi^*\cdot\vec{\nabla}\phi
-\beta(\phi^*\phi-v^2)^2.
\end{equation}
The U(1) symmetry, which is spontaneously broken at
zero temperature, is restored at temperatures above a
critical value $T_c$. Above $T_c$, the field has a
finite correlation length, which diverges at $T_c$. In
the broken phase, the phase angle is a Goldstone mode
and therefore it has an infinite correlation length.

Imagine now that the temperature is initially above
$T_c$, and is gradually decreased so that the system
undergoes a phase transition. When $T_c$ is
approached, the correlation length would have to grow
faster and faster in order to stay in equilibrium.
Sooner or later that becomes impossible. As a result,
the phase angle can only be ordered over some finite
distance $\hat\xi$ in the broken phase, and this leads
to formation of vortices with number density
$n\sim1/\hat\xi^2$~\cite{Kib80}. The length scale $\hat\xi$
has a power-law dependence on the cooling rate, with an
exponent that depends on the universality class of the
transition~\cite{Zur96}.

The spatial vortex distribution can be characterized by 
how the typical net vortex number $\Delta N$
inside a circular curve depends on
its radius $R$. As long as $R\gg\hat\xi$, the curve should pass
through roughly $2\pi R/\hat\xi$ correlated domains, and each time it
moves from one domain to another, the phase changes by a random
amount. As a consequence, the net vortex number should behave like a
random walk of $2\pi R/\hat\xi$ steps,
\begin{equation}
\Delta N(R)\approx \frac{1}{2\pi}\sqrt{\frac{2\pi R}{\hat\xi}}
\propto R^{1/2}.
\label{equ:Kibblerandom}
\end{equation}
For comparison, if the signs of the vortices were randomly
distributed, the net vortex number would grow like $\Delta N(R)\propto
R$. This shows that the vortices formed by this mechanism have
negative correlations.

Superconductors have a spontaneously broken local gauge symmetry,
and therefore 
we modify the Lagrangian by adding
a gauge field, i.e., the electromagnetic vector potential $\vec{A}$,
 \begin{equation}
{\cal L}=\frac{1}{2}\left(\vec{E}^2-B^2\right)
+\dot\phi^*\dot\phi-\vec{D}\phi^*\cdot\vec{D}\phi
-\beta(\phi^*\phi-v^2)^2,
\label{equ:gaugeL}
 \end{equation}
where we have introduced the covariant derivative
$\vec{D}=\vec{\nabla}+ie\vec{A}$, and used the gauge invariance
to set the electric scalar potential to zero. With this choice,
the electric and magnetic fields are given by
$\vec{E}=-\partial\vec{A}/\partial t$ and $B=\partial_1A_2-\partial_2A_1$.

This describes a relativistic, fully two-dimensional superconductor,
and is not a good approximation for a superconductor film, as we will
see later. Nevertheless, we will use it to illustrate the flux
trapping
mechanism of
defect formation~\cite{HinR00}.

Classically, the initial thermal state is given by the Boltzmann
distribution
$p[\phi,A_\mu]\propto\exp(-E[\phi,A_\mu]/T)$, where 
 \begin{eqnarray}
E&=&\int d^2\vec x \left[
\frac{1}{2}\left(\twov{E}^2+B^2\right)
\right.\nonumber\\&&\left.
+\dot\phi^*\dot\phi
+\twov{D}\phi^*\!\cdot\!\twov{D}\phi
+\beta(\phi^*\phi-v^2)^2\right].
 \end{eqnarray}
The most relevant terms for us are the one containing the magnetic
field and the covariant derivative term. 

If the magnetic field
vanishes, the covariant derivative reduces to the ordinary gradient,
and one expects that the above Kibble-Zurek picture is valid.
However, at a non-zero temperature, there are always thermal
fluctuations, which have to be taken into account.
In the symmetric phase, the
effect of the scalar $\phi$ on the magnetic field is
insignificant, and to a good approximation %(in the
%high-temperature limit) 
we can say that $B$ is a
Gaussian random field with a two-point function
 \begin{equation}
\langle B(\twov{x})B(\twov{y})\rangle \approx
T\delta(\twov{x}-\twov{y}),
\end{equation}
or
\begin{equation}
\langle B(\twov{k})B(\twov{q})\rangle \approx
T(2\pi)^2\delta(\twov{k}+\twov{q}).
 \end{equation}
This is simply the classical Rayleigh-Jeans spectrum of blackbody
radiation, and therefore it is only valid when $|\vec{k}|\ll
 k_BT/c\hbar
\approx 440(T/{\rm K}){\rm m}^{-1}$. This constraint may be important
in the case of low-temperature superconductors, but we will not
discuss it here any further..

In the broken phase, the Meissner effect suppresses
the fluctuations, and the two-point function (for
modes with $k\lsim T$) becomes
 \begin{equation}
\langle B(\twov{k})B(\twov{q})\rangle \approx T
\frac{k^2}{k^2+\lambda^{-2}}
(2\pi)^2\delta(\twov{k}+\twov{q}).
 \label{equ:equil2p}
 \end{equation}
where $\lambda$ is the penetration depth,
which starts to decrease from infinity at $T_c$.

However, the magnetic field is conserved,
because $\partial_0B=\twov{\nabla}\cdot \tilde{E}$,
where $\tilde{E}=(-E_2,E_1)$. Therefore, the
long-wavelength modes can only decay very slowly.
Because of causality,  the decay rate must be lower
than the wave number $\gamma(k)\lsim k$, but there are
also other, much stronger constraints arising from the
detailed dynamics. 

In any case, for any cooling rate,
there are going to be some long-wavelength modes that
are too slow to react. Their amplitudes are
therefore frozen to the values they had before the
transition. The corresponding magnetic flux gets trapped into
vortices.
We can define a critical wavenumber
$k_c$, so that this happens to modes with
$k<k_c$. 

Now, consider the typical (root-mean-squared)
magnetic flux through a circle of radius
$R\approx 1/k_c$, 
 \begin{equation}
\Phi(R)=\left\langle\left(\int^{R}
d^2\vec x\,B(\vec x)\right)^2\right\rangle^{1/2}.
 \end{equation}
The whole contribution to this quantity comes from
modes with $k<k_c$, and therefore $\Phi(R)$ does not
change during the transition. After the transition, it
will have the same value as in the symmetric phase
before the transition, and we can calculate that, using
Eq.~(\ref{equ:equil2p}) (and neglecting factors of
order 1),
 \begin{equation}
\Phi(R)\approx \sqrt{T}/k_c.
 \end{equation}
However, because of Meissner effect, this magnetic
flux must be confined into vortices. Each vortex
carries one flux quantum $\Phi_0=2\pi/e$, and
therefore we can say that the typical number of
vortices in a region of radius $R=1/k_c$ is
 \begin{equation}
N(R)=e\Phi(R)\approx
e\sqrt{T}/k_c.
 \end{equation}
Dividing this by the area of such a region gives an
estimate for the number density~\cite{HinR00}
 \begin{equation}
n\approx ek_c\sqrt{T}.
 \end{equation}
Nevertheless, the most obvious difference with the
Kibble mechanism is that all $N(R)$ vortices
formed in any given region have the same sign, which
leads to positive correlations between vortices and
formation of clusters of $N(R)$ vortices.
These clusters have been seen in numerical simulations in
Ref.~\cite{Stephens:2001fv}. 

Note that our discussion was only based on the behaviour of the
long-wavelength magnetic fields, namely the qualitative form of the
two-point function in Eq.~(\ref{equ:equil2p}), causality and the
conservation of magnetic field. It should therefore be independent of
the microscopic details of the system and should apply to real
superconductors as well as to the toy model in Eq.~(\ref{equ:gaugeL}). 
However, as we shall see, one will have to
consider a three-dimensional setup.

\section{Superconducting films}

Superconductor experiments are typically carried out
with a thin superconducting film, but the fully
two-dimensional geometry discussed in
Section~\ref{sect:2d} does not describe that situation
correctly, because the magnetic field should extend
outside the superconductor and induce long-range
interactions between vortices~\cite{Pea64}.

In the symmetric phase, we can calculate the analogue
of  Eq.~(\ref{equ:equil2p}), i.e., the two-point
function on the film from the three-dimensional 
correlation function in the Feynman gauge,
 \begin{equation}
\langle A_i(\threev{k})A_j(\threev{q})\rangle = 
{T\delta_{ij} \over {\bf k}^2}
(2\pi)^3\delta(\threev{k}+\threev{q}).
 \end{equation}
Again, this is equivalent to the classical Rayleigh-Jeans spectrum.
For the magnetic field normal to the film, we find
 \begin{equation}
\langle B_3(\vec x,0) B_3(\vec y, 0)\rangle
=T\int\frac{d^3\threev{k}}{(2\pi)^3}
\frac{\vec k^2}{\threev{k}^2}
e^{i\vec k\cdot(\vec x-\vec y)},
 \end{equation}
where we have set $\threev{k}=(\vec k,k_3)$. 
Equivalently, defining
 \begin{equation}
B(\vec k)=\int d^2\vec x\,B_3(\vec x,0)e^{-i\vec k\cdot\vec x},
 \end{equation}
we have
 \begin{equation}
\langle B(\vec k)B(\vec q)\rangle = 
\frac{Tk}{2}(2\pi)^2\delta(\vec k+\vec q).
 \end{equation}
It is less straightforward to calculate the two-point
function in the broken phase, because the gauge field
is massive on the film but massless outside it.
Therefore, we use a different approach. Deep in the
broken phase, the vortices are well defined, localized
objects, and we can approximate them by point
particles on the film. Because the magnetic field
cannot penetrate the film,  we can flip the direction
of the magnetic field on one side of the film, without
changing the energy of the configuration. This changes
a vortex into a magnetic monopole of charge
$q_M=2\Phi_0=4\pi/e$.  Therefore, the vortices must
behave as magnetic monopoles confined on the film.
Correspondingly, their interaction potential falls off as $1/r$,
as opposed to the $\log(r)$ of global vortices and $\exp(mr)$ of local
vortices in a fully two-dimensional case.
In fact, this result is only valid at distances longer than the Pearl
length $\Lambda_P\approx \lambda^2/d$, where $d$ is the thickness of
the film~\cite{Pea64}. In what follows, we assume that $\Lambda_P$ is
shorter than any of the length scales we discuss.

We now assume that the number density of thermal
vortices is $n_f=n_++n_-$, where $n_+$ and $n_-$ are
the densities of positive and negative vortices,
respectively, and calculate what kind of correlations
the magnetic Coulomb interaction induces. This
calculation is very similar to the calculation of
Debye screening in three dimensions. In the absence of
thermal fluctuations, the magnetic potential around a
vortex would be simply $\phi_0(r)=q_M/4\pi r$. However,
the presence of such a potential biases the thermal
distribution of vortices and antivortices such that
 \begin{equation}
n_\pm(r)=\frac{n_f}{2}\exp(\mp q_M\phi(r)/T)
\approx \frac{n_f}{2}
\left(1\mp \frac{q_M\phi(r)}{T}\right).
 \end{equation}
The magnetic flux density is 
 \begin{equation}
B(r)=q_M(n_+(r)-n_-(r))=-\frac{q_M^2n_f\phi(r)}{T}=
 -\frac{\phi(r)}{\Lambda_{\rm scr}},
 \end{equation}
where we have defined
 \begin{equation}
\Lambda_{\rm scr}=\frac{T}{q_M^2n_f}.
 \label{equ:scrmass}
 \end{equation}
This flux density, in turn, modifies the potential, and we find the
integral equation
 \begin{equation}
\phi({\vec r}) = {q_M\over4\pi r} + 
\int {d^2{\twov{r'}}\,2B({\twov{r'}})\over
4\pi|{\vec r}-{\twov{r'}}|}
= {q_M\over4\pi r} -\frac{2}{  \Lambda_{\rm scr}}
\int {d^2{\twov{r'}}\,\phi({\twov{r'}})\over
4\pi|{\vec r}-{\twov{r'}}|},.
 \end{equation}
We can solve this equation in Fourier space, and find
 \begin{equation}
\phi(\twov{k})=\frac{1}{2}\frac{q_M}{k+\Lambda_{\rm scr}^{-1}}
 \end{equation}
In coordinate space, $\phi$ behaves asymptotically as
 $$
\phi(r)\sim {q_M\over 4\pi r}\quad
 {\rm at}\quad r\ll \Lambda_{\rm scr},
 $$
 $$
\phi(r)\sim
 {q_M\Lambda_{\rm scr}^2\over 4\pi r^3}\quad 
{\rm at}\quad r\gg \Lambda_{\rm scr},
 $$
which shows that the thermal fluctuations screen the
interaction potential, but the potential still has a
power-law form rather than an exponential. Including
the contribution from the original vortex, the
magnetic flux density around a vortex is
 \begin{equation}
B(\twov{r})=\frac{q_M}{2}
\delta(\twov{r})-\frac{\phi(\twov{r})}{\Lambda_{\rm scr}},
 \end{equation}
and since there is a density $n_f$ of these vortices
each with charge $q_M$, the two-point function of the
flux density is
 \begin{equation}
\langle B({\vec x})B({\vec y})\rangle =
 qn_fB(|{\vec x}-{\vec y}|).
 \end{equation}
In Fourier space, this corresponds to
 \begin{eqnarray}
\langle B({\vec k})B({\vec q})\rangle \!\!\!&=&\!\!\! 
qn_f\left[\frac{q_M}{2}-\frac{\phi(k)}{\Lambda_{\rm scr}}\right]
(2\pi)^2\delta(\twov{k}+\twov{q})\nonumber\\
\!\!\!&=&\!\!\!\frac{q_M^2n_f}{2}\frac{k}{k+\Lambda_{\rm scr}^{-1}}
(2\pi)^2\delta(\twov{k}+\twov{q}).
 \end{eqnarray}
We can rewrite this as
 \begin{equation}
\langle B({\vec k})B({\vec q})\rangle = 
\frac{T}{2}\frac{k}{1+k\Lambda_{\rm scr}}
(2\pi)^2\delta(\twov{k}+\twov{q}),
\label{equ:film2p}
 \end{equation}
where we have used Eq.~(\ref{equ:scrmass}).
From this expression, it is easy to see that the
long-distance (low $k$) behaviour agrees with the
symmetric phase, where $\Lambda_{\rm scr}$ is small.

The crucial difference between
Eqs.~(\ref{equ:equil2p})  and (\ref{equ:film2p}) is
that in the case of a film, the important length scale is not the
penetration depth $\lambda$ (or the Pearl length $\Lambda_P$),
but the screening length $\Lambda_{\rm scr}$. 
Whilst $\lambda$ starts from infinity and approaches the zero-temperature
penetration depth, $\Lambda_{\rm scr}$ behaves in the opposite way:
It is initially very short and approaches infinity at zero temperature.

As $\Lambda_{\rm scr}$ increases,
each mode
gets suppressed only when the wavelength becomes shorter than the
screening length, $1/k\lsim \Lambda_{\rm scr}$. Therefore, the
long-wavelength modes feel the effect of the phase
transition later than short-wavelength modes. In fact,
the present case is actually very similar to monopole
formation in three-dimensional non-Abelian
theories~\cite{Raj03},  since the magnetic
charge-charge correlator in that case,
 \begin{equation}
\langle \rho_M(\threev{k})\rho_M(\threev{q})\rangle
\approx T\frac{k^2}{1+k^2/M^2}
(2\pi)^3\delta(\threev{k}+\threev{q}),
 \end{equation}
behaves in the same way.

In any case, Eq.~(\ref{equ:film2p}) shows that if the system could
stay in equilibrium, the fluctuations of all wavelengths would be
suppressed in the zero-temperature limit, where $\Lambda_{\rm
scr}\rightarrow\infty$. 
However, the slowness of the long-wavelength
modes implies that the modes with wave numbers below a
certain critical value $k_c$ freeze out to the
amplitudes they had in the symmetric phase, and we can
approximate
 \begin{equation}
\langle B({\vec k})B({\vec q})\rangle = 
\frac{Tk}{2}\exp(-k/k_c)
(2\pi)^2\delta(\twov{k}+\twov{q}).
 \end{equation}
At this point, we do not specify the value of $k_c$, but only note
 that it must be finite and non-zero. 
The precise form of the cutoff at $k_c$ is irrelevant, and
this exponential form simplifies the calculations,
because  the coordinate space correlator has the
simple form
 \begin{equation}
\langle B({0})B({\vec r})\rangle = 
\frac{Tk_c^3}{4\pi}
\frac{2-r^2k_c^2}{(1+r^2k_c^2)^{5/2}}.
\label{equ:coord2p}
 \end{equation}
This is constant at short distances, $r\ll k_c^{-1}$,
and from that we can deduce the number density
 \begin{equation}
n\approx q_M^{-1}
\sqrt{\lim_{r\rightarrow 0}\langle B(0)B({\vec
r})\rangle} 
=\frac{e}{2\pi}\sqrt{\frac{Tk_c^3}{2\pi}}.
\label{equ:gaugen}
 \end{equation}
This can also be seen by calculating the variance of the
flux $\Phi(R)$ through a circle of radius $R$,
 \begin{eqnarray}
\langle \Phi^2\rangle
&=&\int^R d^2x \int^R d^2y 
\langle B({\vec x})B({\vec y})\rangle
\nonumber\\
&\approx& 
\left\{
\begin{array}{ll}
(\pi/2) T k_c^3 R^4, &\text{for}~R\ll k_c^{-1}\\
TR\ln k_cR,&\text{for}~R\gg k_c^{-1}.
\end{array}\right.
 \label{equ:phi2}
 \end{eqnarray}
At short distances, the typical (rms) flux through a small
curve is proportional to its area and implies a
uniform flux density $B\approx (Tk_c^3)^{1/2}$, in agreement with
 Eq.~(\ref{equ:gaugen}).
At long distances, the scaling is the same as in the Kibble-Zurek
scenario, $\Delta N(R)\sim R^{1/2}$, apart of the logarithm.

By choosing $R=k_c^{-1}$ in Eq.~(\ref{equ:phi2}) we
can also see that the number of vortices in a cluster
is
 \begin{equation}
N_{\rm cl}\approx \sqrt{\frac
{\langle \Phi^2\rangle}{q_M^2}}
\approx \sqrt{\frac{e^2T}{8\pi k_c}}.
 \end{equation}
These results agree with the estimates in
Ref.~\cite{HinR00}.

So far, we have not discussed the value of $k_c$, and indeed, it is
generally more sensitive to the detailed dynamics than the above
considerations. On the other hand, one can carry out an
$k_c$-independent test of the mechanism by measuring, for instance,
the combination $N_{\rm cl}^3n$, which is predicted to be
\begin{equation}
N_{\rm cl}^3n\approx \frac{e^4T^2}{(4\pi)^3}.
\end{equation}

To get a crude estimate of $k_c$, let us assume that Ohm's law
$\vec{j}=\sigma\vec{E}$ is valid, and that the
dynamics of the magnetic field fluctuations is dominated by the
conductivity $\sigma$.
In that case, Maxwell's equations imply that the dynamics of the
magnetic field normal to the film is given by the damped wave equation
\begin{equation}
\frac{d^2B_z}{dt^2}=\partial_z^2 B_z + \vec{\nabla}^2 B_z - \sigma
\frac{dB_z}{dt}.
\end{equation}
Ignoring the $z$ derivative, which would give rise to radiation, we
find that the maximum decay rate of any long-wavelength mode with
wavenumber $|\vec{k}|\ll\sigma$ is
$\gamma(\vec{k})=\vec{k}^2/\sigma$. According to
Eq.~(\ref{equ:film2p}), the amplitude of the mode must decrease as
\begin{equation}
B(\vec{k})\sim \sqrt{T/\Lambda_{\rm scr}}
\end{equation}
for the system to stay in equilibrium
once $\Lambda_{\rm scr}\gsim 1/k$.  The time scale for decrease of
$B(\vec{k})$ is therefore the quench time $\tau_Q$.  The decay
rate $1/\tau_Q$ exceeds the maximum decay rate $\gamma(\vec{k})$, if
\begin{equation}
k<k_c\approx \sqrt{\sigma/\tau_Q}.
\label{equ:kcest}
\end{equation}

\section{Comparison with Experiment}

We now turn to the question of whether these results
can throw any light on recent experimental results. 
Experiments that were very similar to the setup we have discussed
have been carried out by 
Carmi and Polturak~\cite{CarP99} and more recently by Maniv et
al~\cite{Man03}. 
They
involved a rapid quench of a thin film of the
high-temperature superconductor
YBa$_2$Cu$_3$O$_{7-\delta}$, using a SQUID to measure
the net flux following the quench.  This corresponds to the net number
of vortices minus antivortices produced by the quench.

In the first set of experiments~\cite{CarP99}, no evidence for
vortices were found.
The standard causality argument used in cases of
global symmetry breaking \cite{Kib80,Zur96} suggests
that the expected scalar-field correlation length
$\hat\xi$ at the relevant time after the transition
should in this case be about $0.1\,\mu$m.  In the 1\,cm$^2$ area of
the sample, 
the number of vortices $N$ should then be about $10^9$ or
$10^{10}$.    
If one assumes, with Zurek \cite{Zur96},
that the phase of the scalar field makes a random walk
around the perimeter of the sample with a step length
of $\hat\xi$, then one finds $\Delta N\sim N^{1/4}$,
suggesting that $\Delta N$ should be at least of order
100.  The experiment should have seen anything in
excess of $\Delta N\sim20$, so this was thought to be a contradiction.

In the revised experiment~\cite{Man03}, faster cooling rates of around
$dT/dt\approx -10^8$~K/s were
used. This increases the predicted number of vortices, and indeed,
net vortex numbers between 30 and 45 were measured. This number showed
a weak dependence on the cooling rate, with faster quenches producing
more vortices. This is compatible with theoretical predictions.
The Kibble-Zurek scenario with overdamped dynamics would yield a
domain size of $\hat\xi\propto|dT/dt|^{-1/4}$. The net flux would then
be
\begin{equation}
\Phi=\Delta N\Phi_0\approx (R/\hat\xi)^{1/2}\Phi_0\propto |dT/dt|^{1/8},
\label{equ:kibblePhi}
\end{equation}
where $R\approx 1$ cm is the linear size of the film.
The experimental results are consistent with this
scaling~\cite{Man03}, although the error bars are so
large that one cannot use that data to determine the exponent. 
The prefactor in the theoretical prediction is rather uncertain, and
to obtain a fit the authors reduced the prediction by a factor of
four.  Extrapolation to the cooling rate
$20$~K/s used in Ref.~\cite{CarP99} then gives $\Delta N\approx 6$,
which is below the noise level and explains why no vortices were seen
in the original experiment.

However, Carmi and Polturak \cite{CarP99}
rightly pointed out that these arguments are unreliable
in the case of a gauge theory, and suggested an \emph{ad
hoc} modification leading to a prediction $\Delta
N\sim 0.1N^{1/2}\sim10^4$, which is in very severe
disagreement with the experimental results.

We wish to argue, however, that this modification is
not the appropriate way of treating a gauge field.  It
is important to note that the two mechanisms of defect
formation, based on fluctuations of the scalar field
and the gauge field, are characterized by different
length scales.  In the case of the scalar field, it is
the correlation length $\hat\xi$; for the gauge field,
it is $1/k_c$.  So to estimate the contribution to be
expected from magnetic-field fluctuations, we need to
estimate $k_c$.  This is hard to do, because it is
not clear how fast gauge-field correlations can
propagate.  However, we can still put a limit on the
expected number of defects formed.  It seems probable
that $1/k_c$ would be smaller than the size of the
sample.  In that case, the net flux through the
sample should be essentially the same as that in the
high-temperature phase due to thermal fluctuations. 
From
Eq.~(\ref{equ:phi2}), we find
 \begin{equation}
\Delta N \approx \sqrt{e^2TR}.
 \end{equation}
With a sample size $R\approx1$\,cm, this gives $\Delta
N\sim10$.  If in fact $1/k_c$ is larger than $R$, then
the net flux would actually be less than this,
 \begin{equation}
\Delta N \approx \sqrt{e^2Tk_c^3R^4}\propto |dT/dt|^{3/4},
 \end{equation}
where we have used Eq.~(\ref{equ:kcest}).

The above estimate shows that the contribution from the gauge-field 
fluctuations to the net flux is so small that it would have been
below the noise level in Ref.~\cite{CarP99}, just like the
contribution from the Kibble-Zurek mechanism. However, in contrast to
the Kibble-Zurek mechanism, the contribution is independent of the
cooling rate, and the faster cooling used in Ref.~\cite{Man03} would
not help detecting it. At most, it would shift the data points in
Fig.~4 of Ref.~\cite{Man03} up by a small amount, but because the
errors are relatively large and the overall scaling of the theoretical
prediction the data points are compared with is undetermined, one
cannot draw any conclusions about the gauge-field contribution based
on these measurements.

We conclude therefore that it is unlikely that the
magnetic-field fluctutations contribute a detectable
amount to $\Delta N$.  This does not mean, however,
that they can be ignored.  The two mechanisms differ
radically in the nature of the resulting defect
distributions.  If scalar-field fluctuations are
dominant, one expects strong vortex--anti-vortex
correlations --- the two-dimensional analogue of a
preponderance of small vortex loops --- whereas, as
noted above, if magnetic-field fluctuations dominate,
we expect clusters of similarly oriented defects.  

Clearly, it would be very desirable to be able to get
some information on the distribution of vortices, and
in particular to measure $N$ as well as $\Delta N$. 
Only in that way could one separate the contributions
of the two mechanisms.
Achieving this in an experiment with a superconductor film
is probably difficult, because the vortices produced during the quench
may escape the sample in a short time. The flux distribution would
therefore have to be measured very rapidly immediately after the
quench.

However, a recent experiment by Kirtley et al~\cite{Kir03} shows how
this problem can be solved. Instead of a film, they used a square
array of 12$\times$12 thin (50 nm) 
superconducting rings with inner radius of 10
$\mu$m and outer radius of 15 $\mu$m, spaced by 60 $\mu$m.
After the quench, the magnetic flux through any ring can only change
if a vortex can cross the high energy barrier created by the
superconducting ring, and therefore the original fluxes survive a long
enough time to be measured with a scanning SQUID.

For this experiment, the Kibble-Zurek mechanism would predict that each
ring behaves independently of the others. From
Eq.~(\ref{equ:Kibblerandom}), the typical number of
fluxons trapped in a ring would be
\begin{equation}
N\propto |dT/dt|^{1/8}.
\label{equ:Kirtley}
\end{equation}
In fact, there was typically only one flux quantum or no flux at all
through one ring in the experiment, and it is therefore not at all
clear one can use Eq.~(\ref{equ:Kirtley}) to predict the
fluxes. Nevertheless, if one interprets it as the probability of
having a fluxon, the result of the experiment was in a clear
disagreement with it. Instead of a power-law, the measurements showed
a very weak, perhaps logarithmic, dependence on the cooling rate.

Kirtley et al~\cite{Kir03} were able to fit their data with a simple
model, which is similar in spirit to the flux trapping scenario. They
argue that any vortices formed by the Kibble--Zurek mechanism would
be `washed out' by their thermally activated vortex mechanism. 
Because the rings are relatively far apart from each other, they can
be assumed to be independent, and the problem simplifies considerably.
For this reason, the results in Ref.~\cite{Kir03} cannot be thought
to have confirmed the flux trapping scenario. 

On the other hand, it seems that the predictions for the 
spatial vortex distribution, which can best distinguish the two
mechanisms, could be tested with a relatively simple experimental
setup. Instead of a homogeneous film or an array of rings, one would
use a film with small holes in a regular array.  The vortices formed
during a quench would simply move to the nearest hole and get trapped
there. They could then be measured with a scanning SQUID device in
the same way as in Ref.~\cite{Kir03}.

It should be noted, however, that there are many other
complicating factors in the case of a superconducting
film, notably the importance of impurities and pinning
centres, which could drastically affect the vortex
numbers.  Moreover, high-temperature superconductors
involve Cooper pairs with d-wave pairing, so a full
description should involve a multi-component order
parameter.  As a result, the vortices can have more
structure, for example antiferromagnetic cores.  The
effects of these complications on defect formation
remain to be studied.

\acknowledgments

We would like to thank Manuel Donaire, Tom Girard and Ray Rivers for useful
discussions. This work was partly funded by the ESF COSLAB
programme. AR was supported by PPARC and Churchill College, Cambridge.

\end{document}